\documentstyle[preprint,epsf,eqsecnum,aps]{revtex}
\begin{document}
\preprint {Submitted to Phys Rev E, July 4 1996}
\title{Electroconvection in a Suspended Fluid Film:\\ A Linear
Stability Analysis}

\author{Zahir A. Daya and Stephen W. Morris}
\address{Department of Physics and Erindale College\\
University of Toronto\\
60 St. George St.,\\
Toronto, Ontario, Canada M5S 1A7}

\author{John R. de Bruyn}
\address{Department of Physics\\
Memorial University of Newfoundland\\
St. John's, Newfoundland, Canada A1B 3X7}

\date{\today}

\maketitle
\begin{abstract}

A suspended fluid film with two free surfaces convects when a
sufficiently large voltage is applied across it. We present a linear
stability analysis for this system. The forces driving convection are
due to the interaction of the applied electric field with space charge
which develops near the free surfaces. Our analysis is similar to that
for the two-dimensional B\'enard problem, but with important
differences due to coupling between the charge distribution and the
field. We find the neutral stability boundary of a
dimensionless control parameter ${\cal R}$ as a function of the dimensionless
wave number ${\kappa}$. ${\cal R}$, which is proportional to
the square of the applied voltage, is analogous to the Rayleigh number.
The critical values ${{\cal R}_c}$ and
${\kappa_c}$ are found from the minimum of the stability boundary, and
its curvature at the minimum gives the correlation length ${\xi_0}$.
The characteristic time scale ${\tau_0}$, which depends on a second
dimensionless parameter ${\cal P}$, analogous to the Prandtl number, is
determined from the linear growth rate near onset. ${\xi_0}$ and
${\tau_0}$ are coefficients
in the Ginzburg-Landau amplitude equation which describes the flow
pattern near onset in this system.  We compare our results to recent
experiments.

\end{abstract}

\section{Introduction}
\label{introduction}

The regular patterns which form when a dissipative, nonequilibrium
system is driven just beyond the threshold of certain
symmetry-breaking instabilities are in many ways analogous to the
simple ordered phases which appear following equilibrium phase
transitions\cite{revmodphys}.  Patterns can, however, exhibit
interesting nonlinear dynamical behavior, for example chaotic motion,
which has no analog in equilibrium systems. Several fluid-dynamical
systems undergo pattern-forming instabilities which are amenable to
both theoretical and laboratory study.  Examples which have been
extensively explored into the nonlinear regime include Rayleigh-B\'enard
convection\cite{revmodphys,ahlersreview}, Taylor vortex
flow\cite{revmodphys,ahlersreview}, and electroconvection in nematic
liquid crystals\cite{nematicreview}.  In each of these cases, an
essential foundation for understanding the nonlinear behavior is a
complete analysis of the initial linear instability.  The linear
stability analyses for Rayleigh-B\'enard convection and Taylor vortex
flow are classic problems in fluid
mechanics\cite{revmodphys,chandrasekhar,marco}.  The mechanism of the linear
instability for electroconvection in nematic liquid crystals required
many years to elucidate, but even this very complex system is now
reasonably well understood in both the linear and the nonlinear
regimes \cite{nematicreview}.

Our objective in this paper is to carry out a realistic linear
stability analysis for a different electrically driven instability,
namely electroconvection in a thin, suspended fluid film. We have
observed electroconvection patterns in experiments on thin suspended
films of smectic A liquid crystals
\cite{oldprl,jstatphys,PRApaper,endselection,glepaper},
which are isotropic in the plane of the film but have a layered
structure which very strongly impedes flows perpendicular to the film.
As a result, these films can convect rapidly with no change in
thickness. We have observed convection in films only a few
molecules thick.  Our immediate goal is to understand the onset of
electroconvection in this system. The instability mechanism we
describe is, however, not specific to smectic films and would
presumably apply to any sufficiently two-dimensional, weakly
conducting fluid film. In fact, similar convective flows have been
observed in thicker films of nematic and isotropic liquids
\cite{faetti}.  In these cases, however, surface tension effects and
the convective flow itself cause thickness variations in the films,
which make their behavior more complicated than that of the smectic
films.

The model we describe below is physically similar to a highly
simplified one proposed by Faetti, Fronzoni and Rolla\cite{faetti} for
the ``vortex mode" convection they observed in nematic films, but our
analysis is carried much farther.  There is also some similarity
between the driving mechanism considered here and that which drives
electroconvection near the free surface in a partly-filled
capacitor\cite{malkus}.

\begin{figure}
\epsfxsize =4.7in
\centerline{\epsffile{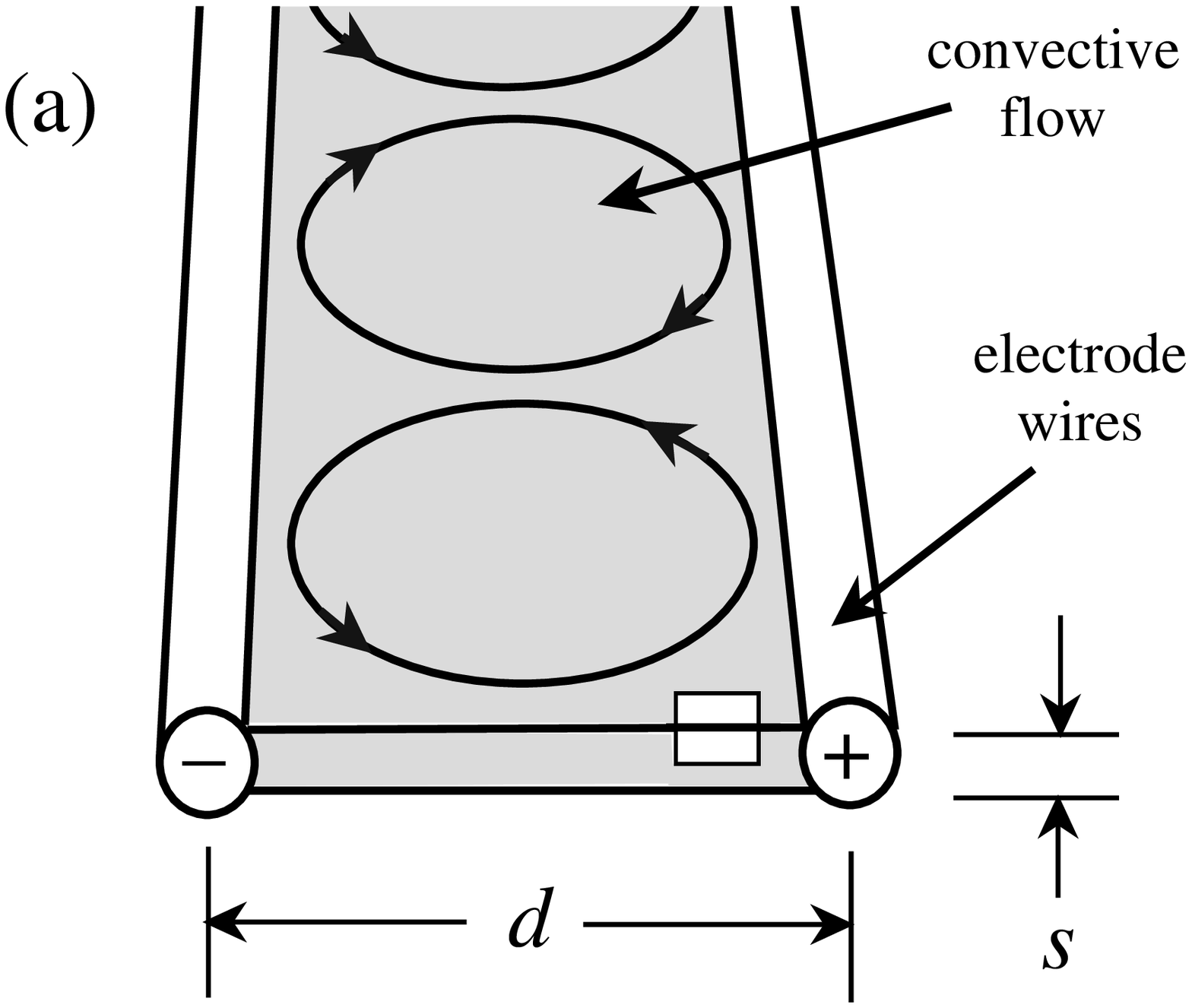}}
\epsfxsize =3.7in
\centerline{\epsffile{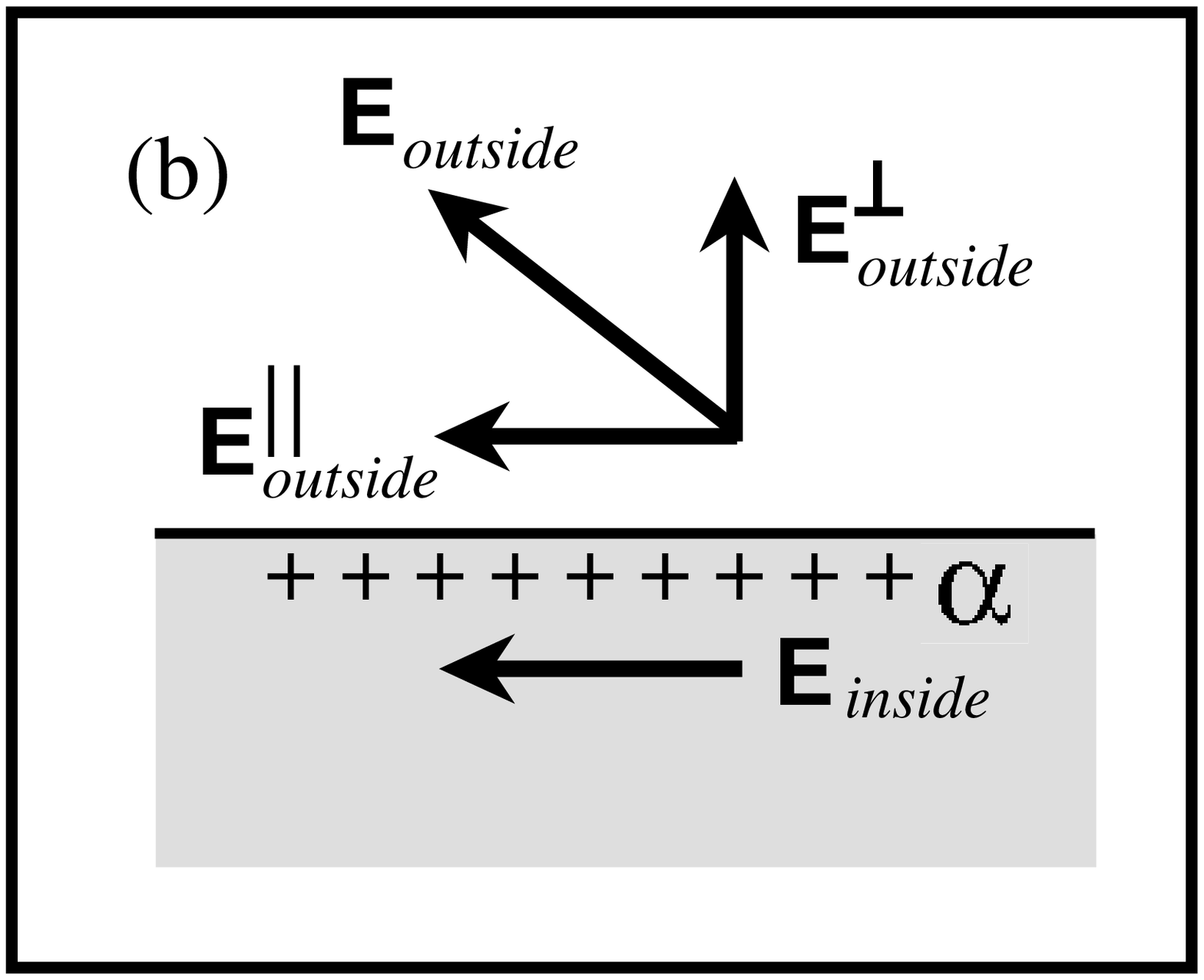}}
\vskip 0.1in
\caption{(a) A schematic of the film and electrodes, as used in
smectic experiments\protect{\cite{PRApaper}}. The film and electrode
are shown enormously exaggerated in thickness; in fact $s/d \approx
10^{-5 }$. (b) Schematic illustration of the fields inside and outside
the film in the small box in part (a).  $\alpha$ is a surface charge
density.  }
\label{3dschematic}
\end{figure}

The relevant experimental arrangement is shown schematically in Fig.
\ref{3dschematic}.  A thin fluid film is suspended between electrodes,
with both its top and bottom surfaces free. The width of the film $d$
is much larger than its thickness $s$, and we will treat it as being
purely two-dimensional.  When the dc voltage applied across the
electrodes exceeds a critical value, the film convects in a pattern of
vortices confined to the plane of the film. We neglect any effects of air
 drag by assuming that the film is suspended in vacuum.  We
will also treat the film as a weak ohmic conductor, and neglect any
electrochemical charge production on the electrodes or in the bulk of
the fluid. The currents involved are assumed to be sufficiently small that
magnetic forces are insignificant.

The body force responsible for driving any electroconvective flow
results from an electric field acting on regions of nonzero charge
density in the fluid. To analyze the electroconvection system we must
first identify the mechanism which gives rise to regions of locally
unbalanced charge in the fluid, and second, solve for the charges and
fields selfconsistently, since these are coupled by Maxwell's
equations. In our model, the charge density arises due
to the electrical boundary conditions at the free surfaces of the
film\cite{PRApaper}.  The inset to Fig. \ref{3dschematic} shows the
essential details
of the charge separation mechanism. Below the onset of convection, the
applied voltage drives a uniform, steady current density $\vec{\bf J}$ through
the film. This is accompanied by a constant electric field $\vec{\bf
E}_{\hbox{inside}} = \vec{\bf J}/\sigma$, where $\sigma$ is the bulk
conductivity. The interior field $\vec{\bf E}_{\hbox{inside}}$ has no
component perpendicular to the film plane.  However, the exterior
field  $\vec{\bf E}_{\hbox{outside}}$ must have both parallel and
perpendicular components just outside the free surfaces of
the film. It cannot in general be
perpendicular to the
surface because the surface is not an equipotential: the film is an
ohmic conductor, so its surface potential varies linearly with the
coordinate between the electrodes.  The parallel component ${\vec{\bf
E}_{\hbox{outside}}}^{||}$ is equal to $\vec{\bf E}_{\hbox{inside}}$
by the usual matching conditions on electric fields across surfaces.
The perpendicular component ${\vec{\bf E}_{\hbox{outside}}}^{\perp}$
is proportional to the surface charge density $\alpha$ at that
location.  It is the interaction of the parallel component of the
field $\vec{\bf E}_{\hbox{inside}} \equiv {\vec{\bf
E}_{\hbox{outside}}}^{||}$ with the surface charge density $\alpha$ at the
two free surfaces that drives the convective flow above the onset of
the instability.

In Section \ref{basestate}, we calculate the surface charge density
below onset by solving for the fields {\it exterior} to the film. This
problem is solved analytically for thin films in two simple electrode
geometries.  We find that a ``charge inversion" is set up in the base state:
the film has a
positive charge density close to the positive electrode, and a
negative charge density close to the negative electrode.  This
inverted charge distribution is sustained by the applied potential
difference across the conductor --- without a potential difference,
the film surfaces are equipotentials, the component of the field parallel
to the surface is zero, and thus there are no forces to drive convection,
even if an electrostatic surface charge is present.  This inverted base
state configuration is analogous to the mass density inversion that
arises in the B\'enard
problem, and in Section \ref{stability} we show that it leads to a
hydrodynamic instablity when the applied voltage is sufficiently
large. Unlike the density in the B\'enard problem, however, the charge
density in the base state is a nonlinear function of position across
the film.

By treating the film as two-dimensional, we neglect the diffusion of
charge on the scale of the film thickness $s$, which acts to smear
the surface charge over a
thickness of order the Debye screening length $\lambda_D$
\cite{PRApaper}. For the very thin films considered here, $\lambda_D$
may be comparable to $s$, in which case the surface
charges and surface forces described above will extend over the whole
thickness of the film. One can show that the total charge contained in
one such Debye layer is the same as that which would reside at one
surface in the absence of diffusion.  The approximation that the film
is a two-dimensional conducting sheet may be expected to break down
for thick films for which $s \gg \lambda_D$. In this limit, surface forces
may lead to significant shears and internal flows, as is apparently
the case in thick nematic and isotropic films \cite{faetti}.
Diffusion on the much larger scale of the film width $d$ is
also neglected.

In Section \ref{stability}, we describe the linear stability analysis
for infinitesimal perturbations about the base state. The stability
calculation is somewhat analogous to that for the
B\'enard problem. Dimensionless quantities ${\cal R}$ and ${\cal P}$
appear, which are analogous to the Rayleigh and Prandtl numbers.
${\cal R}$ is proportional to the square of the applied voltage, while
${\cal P}$ is the ratio of the thin film charge relaxation time to the
viscous relaxation time.

The differences between our calculation and the B\'enard problem are
due to the additional coupling between the field and the charge density,
which is also responsible for the nonlinearity of the base state charge
density.  The charge density and the field are analogous to the
mass density and the gravitational acceleration in the B\'enard problem;
the new requirement that these also satisfy Maxwell's equations amounts
to requiring a nonlocal relation between the charge density and the
electric potential. If we suppress this nonlocality by assuming these
are simply proportional, the base state charge density becomes linear
and our problem reduces completely to the B\'enard case. Interestingly,
this proportionality is nearly correct except near the edges of the film.

In Section \ref{discussion}, we discuss the results for the neutral
curve, and compare the predictions for the critical voltage $V_c$ and
the critical wave number $\kappa_c$ to the values obtained from
experiments. We also calculate the correlation length $\xi_0$ from the
curvature of the neutral curve near $\kappa_c$ and the characteristic
time $\tau_0$ from the linear growth rate at $\kappa_c$.  These
quantities, which are coefficients in the Ginzburg-Landau equation
which describes the amplitude of the pattern near onset, as discussed
below \cite{glepaper,vatche}, are also compared to experimental
results.  Section \ref{conclusion} is a brief summary and conclusion.

\section{The Base State Charge Density}
\label{basestate}

\begin{figure}
\epsfxsize =5in
\centerline{\epsffile{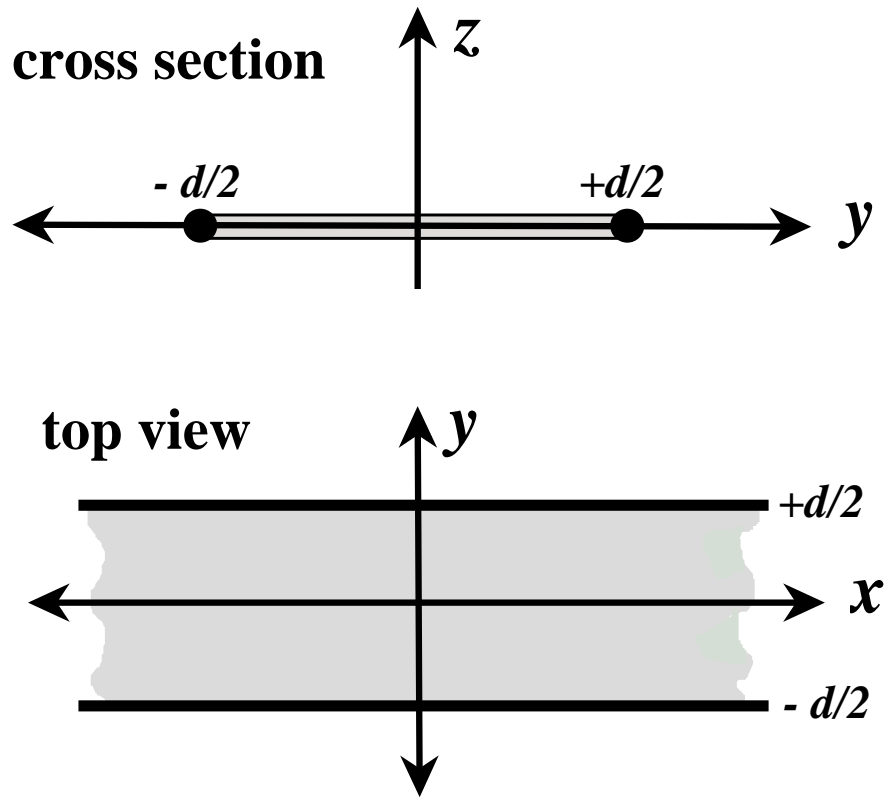}}
\vskip 0.1in
\caption{The coordinates used in the analysis, in which the film is
treated as a two-dimensional sheet.}
\label{coordinates}
\end{figure}

Our first task is to calculate the configuration of charges and
fields below the onset of convection. As described in the previous
section, this is essentially an electrostatic problem in the region
exterior to the film. The coordinates used and the geometry of the
electrodes and film are shown in Fig. \ref {coordinates}.  The origin
is at the centre of the film, which lies between $z=\pm s/2$ and
$y=\pm d/2$.  We will consider the limit of a thin film for which
$s\ll d$. The film is assumed to extend infinitely in the ${x}$
direction.  The upper and lower surfaces of the film are free and the
region outside the film has permittivity $\epsilon_0$. The
permittivity $\epsilon$ of the fluid will turn out to be irrelevant
to the analysis of the base state.

The film will be treated as a charged conducting sheet of negligible
thickness in the $xy$ plane with bulk conductivity $\sigma$.  Its
edges at $y=\pm d/2$ are held at applied potentials $\pm V/2$ by
electrodes of zero thickness.  Below the onset of convection, the film
behaves as an ohmic conductor, so that the potential on the film
varies linearly between $-V/2$ and $+V/2$ for $-d/2 \leq y \leq d/2$.
The potential is zero on the $x$ axis and as $|z| \rightarrow \infty$.
The potential exterior to the film is symmetric above and below the
$xy$ plane and independent of $x$. The charge density $q$ is
proportional to the perpendicular component of the field exterior to
the film, at the film's surface, and so to the $z$ derivative of the
potential there.  To calculate $q$ we need only solve for the
potential in the upper half of the $yz$ plane, subject to boundary
conditions on the $y$ axis. We will consider two simple electrode
geometries, which we refer to as ``plates" and ``wires".  In the plate
geometry, we specify the potential on the rest of the $y$ axis to be
$-V/2$ for $y<-d/2$ and $+V/2$ for $y>d/2$.  Solving for the
potentials is a Dirichlet problem which we solve below using a Green
function. This geometry corresponds to a film held
between infinite knife edges.  Most of the
experiments\cite{oldprl,PRApaper,endselection,glepaper}, however, used
thin wire electrodes to support the edges of the film, as shown
schematically in Fig. \ref{3dschematic}(a). To model this
geometry, the applied potential $\pm V/2$ is specified only at the two
points $y=\pm d/2$.  For $|y|>d/2$ we require that the $z$
derivative of the potential on the $y$ axis be zero.  Thus, in the
upper half of the $yz$ plane, we must solve a mixed boundary value
problem with Dirichlet conditions for $-d/2 \leq y \leq d/2$ and
Neumann conditions for $|y|>d/2$. This is done analytically below,
using the method of dual integral equations\cite{sneddon}.

\subsection{The Base State for Plate Electrodes}
\label{baseplate}

We begin with the simpler plate electrode configuration. We must
solve the Laplace equation for the potential $\Psi$ in the upper
half $yz$ plane,
\begin{equation}
\left(\frac{\partial^2}{\partial y^2} + \frac{\partial^2}{\partial
z^2}\right)\Psi(y,z) = 0,
\label{laplace}
\end{equation}
subject to the piecewise linear Dirichlet boundary conditions on
the $y$ axis
\begin{eqnarray}
 {\Psi(y,0)} & =   -&{V \over 2} \hspace{1cm}  -\infty
<  {y}  \leq -{d \over 2} \nonumber \\
& =&{\frac {V}{d} y}  \hspace{1cm}  -{d \over 2} < {y}  < {d
\over 2} \label{dirichletbc} \\
& = &{V \over 2} \hspace{1.5cm} {d \over 2} \leq  {y}  <
\infty. \nonumber
\end{eqnarray}

\begin{figure}
\epsfxsize =7in
\centerline{\epsffile{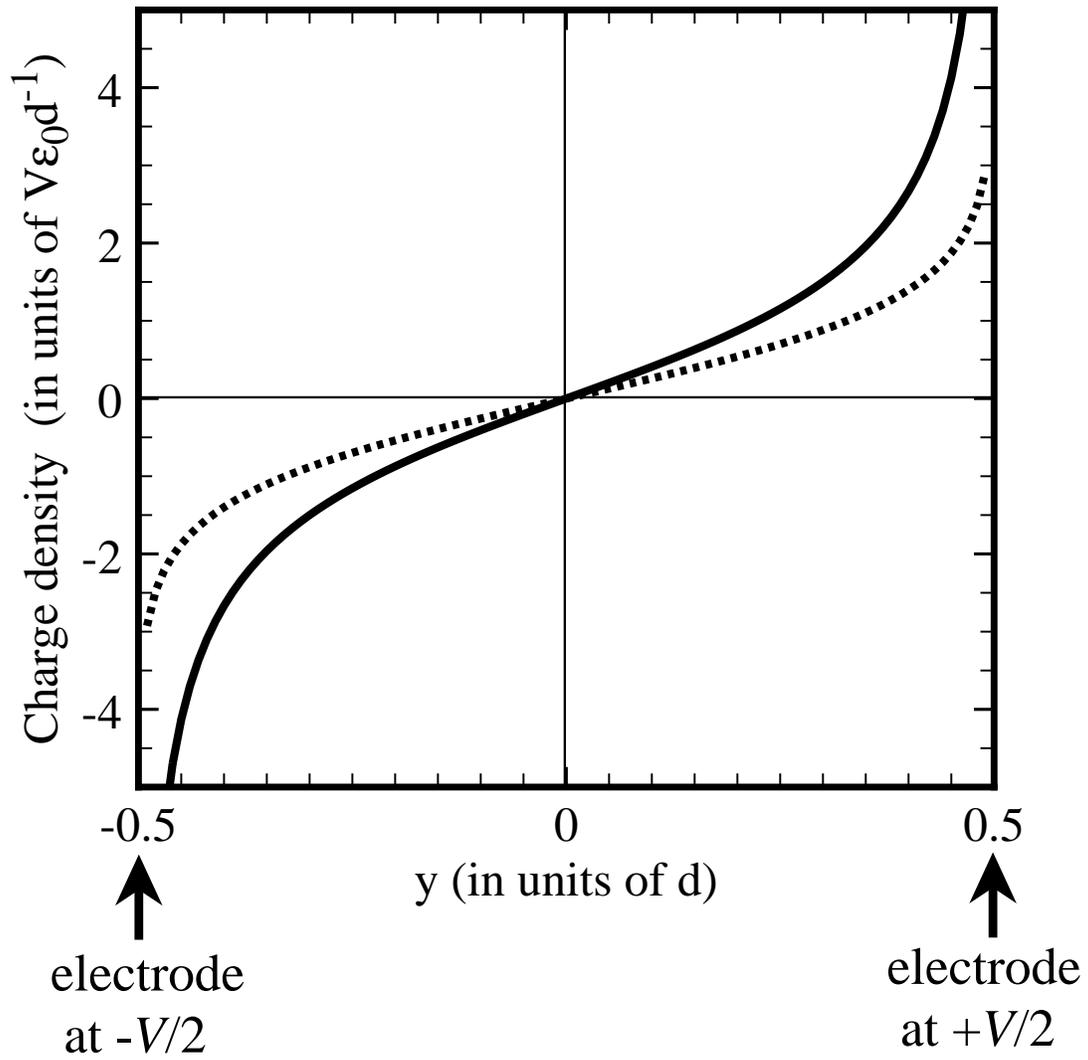}}
\vskip 0.1in
\caption{The surface charge densities for plate (dashed line) and wire
(solid line) electrodes.}
\label{qvsy}
\end{figure}

The appropriate Green function is constructed from a unit line
charge at $(y',z')$ and its image at $(y',-z')$,
\begin{equation}
{G(y,z;y',z')} = {-\log{\frac{(y-y')^2+(z-z')^2}{(y-y')^2+(z+z')^2}}} \>.
\label{greenfun}
\end{equation}
The potential at any point in the upper half plane is given by
\begin{equation}
{\Psi(y,z)} =
{\frac{1}{4\pi}}{\int_{-\infty}^\infty}{\Psi(y',0)}{\frac {\partial
G}{\partial z'}}{\Bigg|_{z'=0}}{dy'} =
{\frac{z}{\pi}}{\int_{-\infty}^\infty}{\frac {\Psi(y',0)}{z^2 +
(y-y')^2}}{dy'} \>.
\label{potgreenfunint}
\end{equation}
The surface charge density on the film is a consequence the fact that
the $z$ components of the electric fields inside and
outside the conducting film are different.  Inside the conductor, in
the absence of diffusion, the $z$ component of the field is
identically zero, as in Fig. \ref{3dschematic}(b), and hence only the
external field is required to determine the surface charge density. As
mentioned in the Introduction, the presence of a diffusion layer near
the surface does not change the total charge density present, per unit
area.  The surface charge density on {\it upper side} of the film is
$-{\epsilon_0}{\partial {\Psi}}/{\partial z}|_{z=s/2}$. In the limit
$s \rightarrow 0$, ${\partial {\Psi}}/{\partial z}$ is discontinuous
across $z=0$, so we use a one-sided derivative valid as $z \rightarrow
0^+$.  To get the total surface charge density $q$ on the film, we
introduce a factor of two to account for the two free surfaces, so
that
\begin{equation}
q = -2 {\epsilon_0} \frac{\partial \Psi}{\partial z}{\Bigg|_{z=0^+}}.
\label{qdefine}
\end{equation}
Using this with Eq. (\ref{potgreenfunint}), interchanging the
order of differentiation and integration, and using Eq.
(\ref{dirichletbc}) gives
\begin{eqnarray}
-{\frac{\pi}{2 \epsilon_{0}} q_{p}(y)}
 =  {\int_{-\infty}^{-\frac{d}{2}}}{\frac{-V/2}{z^2 +
(y-y')^2}}&&{dy'}{\Bigg|_{z=0^+}}
+ {\int_{-\frac{d}{2}}^\frac{d}{2}}{\frac{Vy'/d}{z^2 +
(y-y')^2}}{dy'}{\Bigg|_{z=0^+}}\nonumber \\
+&&{\int_{\frac{d}{2}}^\infty}{\frac{V/2}{z^2 +
(y-y')^2}}{dy'}{\Bigg|_{z=0^+}} \>,
\end{eqnarray}
where the subscript $p$ denotes plate electrodes.  After
integration, the resulting expression was expanded as a power series
in $z$ and evaluated as $z \rightarrow 0^+$.  After some
simplification, we find the surface charge density for the case of
plate electrodes to be given by
\begin{equation}
q_{p}(y)=-{\frac{2V\epsilon_{0}}{\pi
d}}\log{\bigg|}{\frac{y-d/2}{y+d/2}}{\bigg|} \>.
\label{qpdefine}
\end{equation}
This distribution is shown in Fig. \ref{qvsy}. Note that the charge
density is positive near the positive electrode and negative near the
negative electrode, giving the charge inversion described in the
Introduction.  The charge density diverges at the electrodes, which is
an unphysical consequence of the limit $s \rightarrow 0$.  In the real
system, the finite thicknesses of the film and electrodes will impose
a cutoff on $q_{p}$.  This divergence, while unphysical, is weak
enough to be mathematically tractable.  It will turn out that the
rigid boundary conditions we impose on the flow, described in the next
section, ensure that the contributions from the edges of the film are
small. On the other hand, the fact that $q_{p}(y)$ is not a linear
function of $y$ has important consequences for the quantitative
results of the stability analysis.

\subsection{The Base State for Wire Electrodes}
\label{basewire}

We now turn to the case of wire electrodes.  For this case, the
mixed boundary value electrostatic problem can be solved by the
theory of dual integral equations\cite{sneddon}.  We must solve Eq.
(\ref{laplace}), subject to the mixed boundary conditions
\begin{eqnarray}
\Psi(y,0) &=& \frac{V}{d}y\hspace{1cm}\text{for}\qquad |y| \le
d/2 \\
\frac{\partial{\Psi(y,z)}}{\partial z}\bigg|_{z=0^+} &=&
0\hspace{1cm}\text{for} \qquad  |y| > d/2.
\label{mixedbc}
\end{eqnarray}
By separation of variables and using the fact that $\Psi(0,0)=0$, we
make the ansatz that the potential in the upper half plane can be
written as
\begin{equation}
\Psi(y,z)= \int_{0}^\infty\frac{A(k)}{k}e^{-kz}\sin{(ky)}dk.
\label{ansatz}
\end{equation}
With this ansatz, we find the dual integral equations
\begin{eqnarray}
\int_{0}^\infty\frac{A(k)}{k}\sin{(ky)}dk &=&
\frac{V}{d}y\hspace{1cm}\text{for}\qquad |y| < d/2 \\
\int_{0}^\infty A(k)\sin{(ky)}dk &=&0\hspace{1cm}\text{for}
\qquad |y| > d/2.
\label{dual}
\end{eqnarray}
This pair of integral equations may be solved for the potential in
the upper half plane by the method of Sneddon\cite{sneddon}, giving
\begin{equation}
\Psi(y,z)= \frac{V}{2}\int_{0}^\infty\frac{{
J_{1}}(kd/2)}{k}e^{-kz}\sin{(ky)}dk,
\label{potintegral}
\end{equation}
where $J_1$ is the first order Bessel function of the first kind.
Using Eq. (\ref{qdefine}), the total surface charge density is given by
\begin{equation}
{q_{w}(y)} ={ \epsilon_{0} V}\int_{0}^\infty {
J_{1}}(kd/2)\sin{(ky)}dk = \frac{2 \epsilon_{0}
V}{d}\frac{y}{\sqrt{(d/2)^2-y^2}} \>.
\label{qwdefine}
\end{equation}
The subscript $w$ denotes wire electrodes.
This result is also shown in Fig. \ref {qvsy}. As for
the plate electrodes, we find an inverted charge distribution
and divergences at the edges of the film.

Below the onset of convection, the electric field inside the film
is constant and points along ${\bf - \hat y}$.  The force
driving the convection is due to the in-plane electric field acting
on the charge densities $q$.  Unlike the B\'enard problem,
in which the temperature profile is linear below onset, here the
charge density, and hence the body force, is not linear in $y$. This
has the effect of introducing certain non-constant coefficients
into the stability problem, as described in the next section.

\section{Linear Stability Analysis}
\label{stability}

In this Section we consider the stability of the base state to
infinitesimal perturbations. We will show that for sufficiently large
$V$, the electrical forces overcome viscous and conduction losses and
the film becomes unstable to convection.

\subsection{The Perturbation Equations}
\label{perteqns}

Within the thin film, we assume that the fluid velocity $\vec{\bf
u}$ is confined to the film plane, with $\vec{\bf u} = u{\bf \hat x}
+ v{\bf \hat y}$.  In addition, we will neglect any shears in the
$z$ direction.  As discussed above, these assumptions are
reasonable in the context of thin smectic films, where the layer
structure strongly inhibits flow across layers.  We treat the film
as a two-dimensional conducting fluid, with areal material
parameters $\rho_s = s \rho$, $\eta_s = s \eta$, $\sigma_s = s
\sigma$, where $s$ is the film thickness and $\rho$ is the bulk
density, $\eta$ is the bulk molecular viscosity, and $\sigma$ is the
bulk conductivity. (In smectic A films, the viscosity is highly
anisotropic; the relevant component to use for $\eta$ is $\eta_3$,
the viscosity related to shears within layer planes.) The
two-dimensional pressure field is given by $P_s = s P$.  The flow is
driven by the surface force density $q \vec{\bf E}_s$, where
$\vec{\bf E}_s$ is the electric field in the film plane.  The flow
velocity is governed by the two-dimensional Navier-Stokes equation
\begin{equation}
\rho_s \biggl[\frac{\partial \vec{\bf u}}{\partial t} + (\vec{\bf
u} \cdot \nabla_s) \vec{\bf u}\biggr]  =  -\nabla_s P_s + \eta_s
{\nabla_s} ^{2} \vec{\bf u} + q \vec{\bf E}_s,
\label{navierstokes}
\end{equation}
where $\nabla_s$ is the two-dimensional gradient ${{\bf \hat
x}\partial/\partial x + {\bf \hat y}\partial/\partial y}$.
The fluid
is assumed to be incompressible, so that
\begin{equation}
\nabla_s \cdot \vec{\bf u} = 0.
\label{incompress}
\end{equation}
This condition may also be viewed as a constant thickness assumption.
We impose physically realistic rigid boundary conditions on $\vec{\bf
u}$ at the edges of the film, so
\begin{equation}
\vec{\bf u} \equiv 0 \quad\text{ and }\quad  \frac{\partial v}{\partial
y} \equiv 0 \quad\text{ at }\quad y = \pm d/2.
\label{rigidbc}
\end{equation}
The motion of charge is governed by the charge continuity equation
\begin{equation}
\frac{\partial q}{\partial t}  = -\nabla_s \cdot {\vec{\bf J}_s} =
-\nabla_s \cdot (q\vec{\bf u} + \sigma_s {\vec{\bf E}_s}),
\label{chgcontinuity}
\end{equation}
in which ${\vec{\bf J}_s}$ is the two-dimensional current density
in the plane of the film, and includes contributions from both
conduction ($\sigma_s {\vec{\bf E}_s}$) and convection ($q\vec{\bf
u}$). Diffusion of charge in the plane of the film has been neglected.

The electric
field in the plane of the film ${\vec{\bf E}_s}$ is given by
\begin{equation}
{\vec{\bf E}_s} = - \nabla_s \Psi_s = -\nabla_s \Psi \big|_{z=0} \>.
\label{surfE}
\end{equation}
As in the previous section, $\Psi$ is the potential which solves
the {\it three-dimensional\/} Laplace equation,
\begin{equation}
\nabla^2 \Psi = 0,
\label{3dlaplace}
\end{equation}
in the half space $z \geq 0$, with the surface charge density given by
Eq. (\ref{qdefine}).

The surface charge density $q$ in Eqs. (\ref{navierstokes}) and
(\ref{chgcontinuity}) in principle contains both the density of
free charges $q_{\hbox{free}}$ and that of the dielectric polarization
charges, so that $q = q_{\hbox{free}} - \nabla_s \cdot \vec{\bf P}_s$.
Here $\vec{\bf P}_s = s \vec{\bf P}$, where $\vec{\bf P}$, the bulk
polarization density, is given by  $\vec{\bf P} = \epsilon_0
\chi \vec{\bf E}$, where $\chi$ is the electric susceptibility.
Inside the film, ${\vec{\bf E}} = {\vec{\bf E}_s}$, independent
of $z$, and has zero $z$ component. Eq. (\ref{qdefine}),
which only involves exterior fields, therefore only gives
$q_{\hbox{free}}$. In the base state, however, $\nabla_s \cdot
\vec{\bf E}_s = 0$ and $q = q_{\hbox{free}}$, so polarization
effects are irrelevant. In the general case, one can show using
the scalings given below that the dimensionless form of the
$\nabla_s \cdot \vec{\bf P}_s$ terms are proportional to $\chi s
/ d$. For the experiments of interest, $\chi s / d \approx 10^{-4}$,
 so we can safely neglect polarization effects.

Eqs. (\ref{navierstokes})-(\ref{3dlaplace}) have a simple solution when
$\vec{\bf u} \equiv 0$.  Eq. (\ref{3dlaplace}), subject to the
appropriate boundary conditions, corresponds to the base state charge
density problem solved in the previous section.  Once the fields and
$q$ are found from Eqs. (\ref{qdefine}) and (\ref{3dlaplace}), Eq.
(\ref{navierstokes}) can be solved for the pressure $P_s$ whose gradient
balances the surface force density $q \vec{\bf E}_s$. Eq.
(\ref{chgcontinuity}) then gives a constant current density ${\vec{\bf
J}_s} = \sigma_s {\vec{\bf E}_s}$, with a constant interior field
${\vec{\bf E}_s} = -(V/d) \hat{\bf y}$.  This is a selfconsistent
solution which corresponds to hydrostatic
equilibrium. The pressure gradient points everywhere toward the
midline of the film, $y=0$.

Expanding the divergence in Eq. (\ref{chgcontinuity}) and using Eq.
(\ref{incompress}), the charge continuity equation becomes
\begin{equation}
\frac{\partial q}{\partial t} = - \vec{\bf u} \cdot \nabla_s q - \sigma_s
\nabla_s \cdot {\vec{\bf E}_s} \>.
\label{chgcont2}
\end{equation}
Note that $\nabla_s \cdot {\vec{\bf E}_s}$ is not equivalent to
$\nabla \cdot {\vec{\bf E}}$, because of discontinuities in the $z$
component of ${\vec{\bf E}}$ at $z=0$.  Thus, $\nabla_s \cdot
{\vec{\bf E}_s}$ is not directly related to the charge density $q$.
$\nabla_s \cdot {\vec{\bf E}_s}$ can be found only after solving the
full three-dimensional Laplace problem given by Eq. (\ref{3dlaplace}).

To examine the stability of the base state, we introduce perturbed quantities
\begin{eqnarray}
\vec{\bf u} & = & 0 + \vec{\bf u}^{(1)}, \\
q & = & q^{(0)} + q^{(1)}, \\
P_s & = & {P_s}^{(0)} + {P_s}^{(1)}, \\
 \vec{\bf E}_s& = & \vec{\bf E}_s^{(0)} + \vec{\bf E}_s^{(1)}.
\label{perturbs}
\end{eqnarray}
where $\vec{\bf E}_s^{(0)} = {E_y}^{(0)}{\bf \hat y}$ and $\vec{\bf
E}_s^{(1)}={E_{x}}^{(1)}{\bf \hat x}+{E_{y}}^{(1)}{\bf \hat y}$.
Here ${E_y}^{(0)} = -V/d$, and $q^{(0)}$ is the base
state charge density found in the previous section. To first order
in the small perturbations, Eqs.
(\ref{navierstokes}), (\ref{incompress}), and (\ref{chgcont2})
become
\begin{eqnarray}
\nabla_s \cdot \vec{\bf u}^{(1)} & = & 0  \\
\rho_s \frac{\partial \vec{\bf u}^{(1)}}{\partial t} & = & -\nabla_s
{P_s}^{(1)} + \eta_s {\nabla_s} ^{2} \vec{\bf u}^{(1)} +
q^{(0)}{E_{x}}^{(1)}\hat{\bf x}+ q^{(1)} {E_y}^{(0)} \hat{\bf y} +
q^{(0)}{E_{y}}^{(1)}\hat{\bf y}  \label{ns1} \\
\frac{\partial q^{(1)}}{\partial t} & = & - \vec{\bf u}^{(1)} \cdot
\nabla_s q -  \sigma_s\biggl[\nabla_s \cdot ({E_{x}}^{(1)}\hat{\bf
x}+{E_{y}}^{(1)}\hat{\bf y}) \biggr]. \label{conseqzz}
\end{eqnarray}
Taking the curl of Eq. (\ref{ns1}) eliminates the pressure.  Taking a
second curl and using Eq. (\ref{incompress}) gives
\begin{equation}
\rho_s \frac{\partial}{\partial t}  {\nabla_s} ^{2} \vec{\bf
u}^{(1)}  =  \eta_s {\nabla_s} ^{2} {\nabla_s} ^{2} \vec{\bf
u}^{(1)} - \nabla_s \times \biggl[ \nabla_s \times
\bigl(q^{(0)}{E_{x}}^{(1)}\hat{\bf x} + q^{(1)} {E_y}^{(0)} \hat{\bf
y} + q^{(0)}{E_{y}}^{(1)} \hat{\bf y}  \bigr)\biggr].
\label{ns2}
\end{equation}
{}From this equation, we select the $y$ component, which is
\begin{equation}
\rho_s \frac{\partial}{\partial t}  {\nabla_s} ^{2} v^{(1)} =
\eta_s {\nabla_s} ^{2} {\nabla_s} ^{2} v^{(1)} + {E_y}^{(0)}
\frac{\partial ^{2} q^{(1)}}{\partial x^{2}} - \frac{\partial
q^{(0)}}{\partial y}\frac{\partial {E_{x}}^{(1)}}{\partial x}
\>.
\label{ns2y}
\end{equation}
Using the fact that $q^{(0)}$ is only a function of $y$,
Eq. (\ref{conseqzz}) becomes
\begin{equation}
\frac{\partial q^{(1)}}{\partial t} =
- v^{(1)} \frac{\partial q^{(0)}}{\partial y} - \sigma_s
\biggl(\frac{\partial{E_{x}}^{(1)}}{\partial
x}+\frac{\partial{E_{y}}^{(1)}}{\partial y}\biggr).
\label{chgcont3}
\end{equation}
In the previous equations, $\vec{\bf E}_s^{(1)}$ is the in-plane
electric field produced by the charge distribution $q^{(1)}$.
Introducing perturbations for the electric potential with ${\Psi}=
{\Psi}^{(0)} + {\Psi}^{(1)}$, we have ${\vec{\bf E}_s}^{(1)} = -
\nabla_s {\Psi_s}^{(1)} = -\nabla_s \Psi^{(1)} \big|_{z=0}$. The
perturbation potential $\Psi^{(1)}$ is the solution of a new
three-dimensional Laplace problem in $z \geq 0$ analogous to Eq.
(\ref{3dlaplace}),
\begin{equation}
\nabla^2 \Psi^{(1)} = 0,
\label{3dlaplacepert}
\end{equation}
with $q^{(1)}=-2 {\epsilon_0} (\partial \Psi^{(1)}/ \partial z
)|_{z=0^+}$.

We now replace the various field components with the appropriate
derivatives of the potential in Eqs. (\ref {ns2y}) and (\ref {chgcont3}), to
get
\begin{equation}
  {\rho_s} \frac{\partial}{\partial t} {\nabla_s} ^{2}
v^{(1)} =
{\eta_s} {\nabla_s} ^{2} {\nabla_s} ^{2} v^{(1)} + {E_y}^{(0)}
\frac{\partial ^{2} {q}^{(1)}}{\partial x^{2}} + \frac{\partial
{q}^{(0)}}{\partial y}\frac{\partial^2 {\Psi}^{(1)}_s }{\partial
x^2}
\label{ns3y}
\end{equation}
\begin{equation}
\frac{\partial {q}^{(1)}}{\partial t} =
- v^{(1)} \frac{\partial {q}^{(0)}}{\partial y} + {\sigma_s}
\biggl(\frac{\partial^2 {\Psi}^{(1)}_s}{\partial
x^2}+\frac{\partial^2 {\Psi}^{(1)}_s}{\partial y^2}\biggr) \>.
\label{chgcont4}
\end{equation}
Eqs. (\ref{3dlaplacepert})-(\ref{chgcont4}) are the
equations for the perturbations that we must solve to
determine the stability of the base state.

The specification of the boundary conditions necessary to solve Eq.
(\ref{3dlaplacepert}) requires some explanation.  By writing ${\Psi}=
{\Psi}^{(0)} + {\Psi}^{(1)}$, we split the full Laplace problem of Eq.
(\ref{3dlaplace}) into separate Laplace problems at each order.  At
zeroth order, the boundary conditions at $z=0$ on ${\Psi}^{(0)}$ are
those described in Section \ref{basestate} for each electrode
configuration in the base state. In particular, ${\Psi}^{(0)}$ was set
equal to $\pm V/2$ at the edges of the film.  At first order, the
boundary conditions on ${\Psi}^{(1)}$ require that ${\Psi}^{(1)} = 0$
at the edges of the film, and on both of the electrodes in the plate
case. In the wire electrode case we require $\partial
{\Psi}^{(1)}/\partial z |_{z=0^+} = 0$ for $|y| > d/2$ and $z=0$. In
either electrode case, we will find Dirichlet boundary conditions for
${\Psi}^{(1)}$ on the film itself by selfconsistently solving Eq.
(\ref{chgcont4}).  Proceeding in this way, the boundary conditions on
the total potential ${\Psi}$ are satisified by the superposition of
${\Psi}^{(0)}$ and ${\Psi}^{(1)}$.

\subsection{The Normal Mode Expansion}
\label{modeexpansion}

We now expand the velocity, charge density and potential
perturbations in normal modes which are periodic in $x$ with
wavenumber $k$, and have growth rate $\gamma$,
\begin{eqnarray}
v^{(1)} & = & \Lambda(y) e^{ikx+\gamma t}\>, \label{aaa}\\
{q}^{(1)} & = & \Theta(y, k,\gamma)e^{ikx+\gamma t}\>, \label{bbb} \\
{\Psi}^{(1)}& = & \Omega(y, z, k,\gamma) e^{ikx+\gamma t}\>.\label{ccc}
\label{normalmodes}
\end{eqnarray}
We substitute Eqs. (\ref{aaa})-(\ref{ccc}) into Eqs.
(\ref{3dlaplacepert})-(\ref {chgcont4}) and non-dimensionalize the system
by dividing lengths by $d$, times by $\epsilon_{0}d/\sigma_s$ and
charge densities by $\epsilon_{0}V/d$.  We then write
$D={\partial}/{\partial y}$ and define new dimensionless quantities
$\kappa =kd$ and $Q(y)= d^2 D{q}^{(0)}(y) /\epsilon_{0}V$. The
resulting dimensionless equations are
\begin{equation}
(D^2-{\kappa}^2)\biggl(D^2-{\kappa}^2-\frac{\gamma}{{\cal
P}}\biggr)\Lambda+{\kappa}^2 {\cal R}
\bigl(\Theta-Q\Omega_s\bigr)=0
\label{ndnslin}
\end{equation}
and
\begin{equation}
(D^2-{\kappa}^2)\Omega_s -\gamma\Theta-Q\Lambda=0,
\label{ndchcontlin}
\end{equation}
where $\Omega_s = \Omega|_{z=0}$. The three-dimensional Laplace
equation, Eq. (\ref{3dlaplacepert}), becomes
\begin{equation}
{\nabla}^{2}(\Omega e^{i\kappa x}) =
\Biggl[\frac{\partial^2}{\partial y^2}+\frac{\partial^2}{\partial
z^2}-{\kappa}^2\Biggr]\Omega= 0,
\label{Helmholtz}
\end{equation}
with Eq. (\ref{qdefine}) imposing the condition that
\begin{equation}
\Theta = -2\frac{\partial \Omega}{\partial z}\Bigg|_{z=0^+}~.
\label{theta}
\end{equation}
Eq. (\ref{Helmholtz}) is a {\it two-dimensional\/} Helmholtz equation
in the half plane $x=0$, $z \geq 0$, which is perpendicular to the
plane of the film. Eqs. (\ref{Helmholtz}) and (\ref{theta}) determine the
rather complicated nonlocal coupling between the in-plane potential
function $\Omega_s(y,\kappa,\gamma) = \Omega(y,0,\kappa,\gamma)$
and the charge density function $\Theta(y,\kappa,\gamma)$.

Two dimensionless groups appear: ${\cal R}$, which plays the part
of the Rayleigh number, and ${\cal P}$, which plays the part of the
Prandtl number.  In terms of the bulk, rather than surface, material
parameters, they are given by
\begin{equation}
{\cal R} = \frac{{\epsilon_0}^2 V^2}{\sigma \eta s^2}
\label{Rdefine}
\end{equation}
and
\begin{equation}
{\cal P} = \frac{\epsilon_0 \eta}{\rho \sigma d s}\>.
\label{Pdefine}
\end{equation}
${\cal R}$, the control parameter, is proportional to $V^2$.  It is
interesting to note that ${\cal R}$ is independent of $d$, the
width of the film.  The Prandtl-like parameter ${\cal P}$ may be
regarded as the ratio $\tau_q/\tau_v$ of the two time scales in the
problem, the charge relaxation time for thin films\cite{PRApaper} $\tau_q =
\epsilon_0 d / \sigma s$ and the viscous relaxation
time $\tau_v = \rho d^2 / \eta$.

The effect of the nonlinear $y$ dependence of the derivative of the
base state charge density $D{q}^{(0)}(y)$ is contained in the
non-constant coefficient $Q(y)$. For the plate and wire electrode
configurations, we find from Eqs.
(\ref{qpdefine}) and (\ref{qwdefine}) that $Q(y)$ is given
by
\begin{equation}
{Q_{p}}(y)  =
\frac{8}{\pi{(1-4y^2)}}
\label{Qplate}
\end{equation}
and
\begin{equation}
{{Q}_{w}}(y)  =
\frac{4}{(1-4y^2)^{3/2}} \>,
\label {Qwire}
\end{equation}
respectively.

\subsection{Analogy to the B\'enard Problem}
\label{benardanalogy}

The above equations bear a strong analogy to the corresponding equations
in the B\'enard problem.  The correspondence becomes complete if the
nonlocal coupling of the charges and potentials given by Eq.
(\ref{Helmholtz}) and (\ref{theta}), is suppressed by simply putting $q
\propto \Psi_s$. Applying this assumption to the base state removes the
nonlinear $y$ dependence of the charge density so that $Q(y) \equiv 1$. In
fact, detailed analysis shows that $q$ is always nearly proportional to
$\Psi_s$ in the central part of the film. This can be seen, for example,
in Fig. \ref{qvsy} near $y=0$. If this proportionality is assumed to hold
everywhere, then our continuity equation for charge, Eq.
(\ref{chgcontinuity}) becomes identical to the thermal diffusion equation
in the B\'enard problem.  Under the same assumption, the force term $q
{\vec{\bf E}_s}$ in Eq. (\ref{navierstokes}) becomes proportional to $q
\hat{\bf y}$, which is the form of the analgous term in the B\'enard
problem.  Turning the argument around, one can say that the reason that
our system {\it does not} reduce to the B\'enard problem is because the
charges and fields are nonlocally coupled {\it via} the the charge
distribution's own self-field.

\subsection{The Compatibility Condition}
\label{compatibility}

To find the conditions which ${\cal R}$, ${\cal P}$, $\kappa$ and
$\gamma$ must satisfy for solutions to exist, we solve the
linearized equations by means of various expansions. A crucial step
that must be done numerically is the solution of the Helmholtz
equation in the plane perpendicular to the film,  Eq.  (\ref{Helmholtz}),
which necessarily involves a numerical relaxation calculation.

At the edges of the film, $y=\pm 1/2$, the rigid boundary
conditions on the flow velocity $\vec{\bf u}^{(1)}$, given by Eq.
(\ref{rigidbc}), require that $\Lambda(y)$ satisfy the four conditions
\begin{equation}
\Lambda(\pm 1/2)= D\Lambda(\pm 1/2) =0.
\label{lambdaBC}
\end{equation}
To ensure this, we expand $\Lambda(y)$ as
\begin{equation}
\Lambda(y) = \sum\limits_{m=1}^\infty A_m C_m(y),
\label {lamdaexpandchan}
\end{equation}
where the $C_m(y)$ are even Chandrasekhar
functions\cite{chandrasekhar},
\begin{equation}
C_m(y) = \frac{\cosh(\lambda_m y)}{\cosh(\lambda_m /2)}-
\frac{\cos(\lambda_m y)}{\cos(\lambda_m /2)} \>.
\label{Cmdefine}
\end{equation}
Here $\lambda_m$ is the $m$th root of $\tanh(\lambda_m
/2)+\tan(\lambda_m /2)=0$ \cite{oops} .
We can restrict the expansion to even functions because
of the symmetry of the equations about $y=0$. Note that an expansion
in $C_m$ has been shown to give a good description of the velocity
field measured in experiments on electroconvection in smectic films
\cite{PRApaper}.  Only relative amplitudes matter in Eq.
(\ref{lamdaexpandchan}), so we set $A_1=1$. It follows from linearity
that we can also write $\Omega=\sum_m A_m{\Omega}_m$ and $\Theta =
\sum_m A_m{\Theta}_{m}$, where ${\Omega}_m$ and $\Theta_{m}$ are the
solutions corresponding to $\Lambda=C_m$. As above, we denote
$\Omega_{sm} = \Omega_{m}|_{z=0}$.

\subsubsection{The potential function \lowercase{$\Omega_{sm}$} for
\lowercase{$\gamma=0$}}

Setting $\gamma = 0$ and substituting $C_m$ for $\Lambda$ in Eq.
(\ref{ndchcontlin}) gives
\begin{equation}
(D^2-{\kappa}^2)\Omega_{sm} = Q C_{m},
\label{omega}
\end {equation}
which may be solved directly by Fourier expansion. Since $Q C_{m}$
is even, we expand
\begin{equation}
Q C_{m}=\sum_{n=0}^{\infty} b_{mn} \cos{(2 n \pi y)},
\label{QCexpand}
\end{equation}
in which
\begin{equation}
b_{m0}=2\int_{0}^{\frac{1}{2}} Q(y) C_{m}(y)
dy
\label{bm0}
\end{equation}
and
\begin{equation}
b_{mn}=4\int_{0}^{\frac{1}{2}} Q(y) C_{m}(y)
\cos{(2n\pi y)} dy\>.
\label{fouriercoef}
\end{equation}
Using a similar Fourier expansion of $\Omega_{sm}$ and imposing
the zero boundary conditions at $y= \pm 1/2$, we find
\begin{equation}
{\Omega}_{sm} =\sum_{n=0}^{\infty} \frac{b_{mn}}{[(2n \pi)^{2} +
{\kappa}^{2}]} \Biggl[\frac{(-1)^{n}\cosh{(\kappa y)}}{\cosh{(\kappa
/2)}} - \cos{(2n \pi y)}\Biggr] \>.
\label{omegam}
\end{equation}
To calculate $\Omega_{sm}$, we used a Romberg numerical integration
scheme\cite{recipes} to tabulate the integrals for $b_{mn}$ in Eqs.
(\ref{bm0}) and (\ref{fouriercoef}) for the each of the two electrode
geometries, using $Q_p$ and $Q_w$ as given by Eqs. (\ref{Qplate}) and
(\ref{Qwire}). We used an upper cutoff of $n=29$, which was dictated by
the double precision accuracy of the Romberg scheme.

\subsubsection{The potential function \lowercase{$\Omega_{sm}$} for
\lowercase{$\gamma \neq 0$}}
\label{gammanonzero}

For nonzero $\gamma$, we solved
\begin{equation}
(D^2-{\kappa}^2)\Omega_{sm} = Q C_{m} + \gamma \Theta_m
\label{omegagamma}
\end {equation}
by an iterative scheme.  We used the $\gamma=0$ solution, Eq.
(\ref{omegam}), to find a first approximation ${\Omega_{sm}}^{[0]}$.
{}From this, we calculated the corresponding approximate the charge
density function ${\Theta_m}^{[0]}$ using the relaxation algorithm
described in Section \ref{relax} below.  Then $Q C_{m} + \gamma
{\Theta_m}^{[0]}$ was Fourier expanded in the same manner as $Q C_{m}$
in Eqs. (\ref{QCexpand})-(\ref{fouriercoef}) above.  This
expansion was used to find a series solution analogous to Eq.
(\ref{omegam}) for the next approximation ${\Omega_{sm}}^{[1]}$, which
was then relaxed to find ${\Theta_m}^{[1]}$.  This sequence of
steps was iterated until it converged for both ${\Omega_{sm}}$ and
${\Theta_{m}}$.  The convergence criterion was a sum of the squares
of 100 differences in successive iterates distributed on $0 \leq y
\leq 1/2$. For $|\gamma| \leq 0.1$, the sum converged after 7 or 8
iterations to a precision limited by the Romberg integration scheme
used to find the Fourier coefficients.

\subsubsection{The charge density function \lowercase{$\Theta_{m}$} }
\label{relax}

We solved the Helmholtz equation, Eq. (\ref{Helmholtz}), for $\Omega_m$
for each of the two electrode geometries, using a simple SOR
algorithm\cite{recipes}.  In each case, the Dirichlet conditions on
$\Omega_m$ for $-1/2 \leq y \leq 1/2$ and $z=0$ are given by
${\Omega}_{sm}$ (Eq. (\ref{omegam})) in the case of $\gamma=0$, or by
the corresponding expression during iteration for $\gamma
\neq 0$. Beyond the film, for $|y|>1/2$, $z=0$, we applied the
Dirichlet condition $\Omega_m=0$ in the plate electrode case, and
Neumann conditions $(\partial \Omega_m/\partial z)|_{z=0}=0$ in the
wire case.

Because $\Omega_m$ is even in $y$, it need only be relaxed in the
first quadrant of the $yz$ plane. We used an $N \times N$ square
lattice of cells in this quadrant, with $N_{film} < N$ points
between $y=0$ and $y=1/2$. On the outer edges of the lattice, we set
$\Omega_m=0$ to enforce the zero boundary condition at infinity.
Starting with $N=100$ and $N_{film}=50$, we systematically increased
$N$ and $N_{film}$ is such a way that $N_{film}/N \rightarrow 0$.
All the quantities calculated below showed a small residual
monotonic variation with $N_{film}$; we removed this by plotting
each against $1/N_{film}$ and extrapolating to $1/N_{film}
\rightarrow 0$.

{}From the resulting $\Omega_m$, the charge density perturbation
$\Theta_m$ was determined from Eq. (\ref{theta}) by taking the
one-sided $z$ derivative numerically.  $\Theta_m$ was therefore only
known at $N_{film}$ lattice points across the film.  For the
purposes of integration, we used a Chebyshev
interpolation\cite{recipes} of these points.

\subsubsection{The compatibility condition}

To find the general compatibility conditions on solutions, we
substitute the expansions for $\Lambda$, $\Omega_s$ and $\Theta$
into Eq. (\ref {ndnslin}) to get
\begin{equation}
\sum\limits_{m=1}^\infty \Biggl[ (D^2-{\kappa}^2)\biggl(D^2-
{\kappa}^2-\frac{\gamma}{{\cal
P}}\biggr)C_m
+{\kappa}^2{\cal R}\bigg({\Theta_m}-Q{{\Omega}_{sm}}\biggr)\Biggr]A_m =0.
\label{homog}
\end{equation}
Multiplying by the Chandrasekhar function $C_l(y)$ and integrating
from $y=-1/2$ to $y=+1/2$, we form inner products, denoted by
$\langle \cdot\cdot\cdot \rangle$. Then Eq. (\ref{homog}) becomes a
linear homogenous system with the determinant compatibility
condition
\begin{equation}
\Biggl{\|}\left\langle
C_l(D^2-{\kappa}^2)\biggl(D^2-{\kappa}^2-\frac{\gamma}{{\cal
P}}\biggr)C_m \right\rangle \\
+{\kappa}^2{\cal R}\bigg\langle
C_l\Big({\Theta_m}-Q{{\Omega}_{sm}}\Big) \bigg\rangle\Biggr{\|}=0.
\label{compatlong}
\end{equation}
After some simplification, this can be written as
\begin{equation}
\Biggl{\|}(\lambda_{m}^{4} + {\kappa}^4)\delta_{lm} - 2{\kappa}^2
X_{lm} + {\kappa}^2 {\cal R} F_{lm} -\frac{\gamma}{{\cal P}}
(X_{lm}-{\kappa}^2 \delta_{lm})\Biggr{\|}=0,
\label{compat}
\end{equation}
where $F_{lm}=\langle C_{l}(\Theta_{m}-Q{{\Omega}_{sm}}) \rangle$.
The matrix elements $X_{lm}$ are given analytically
by\cite{chandrasekhar}
\begin{eqnarray}
X_{lm}&=& \langle C_{l}^{\prime\prime} C_{m}\rangle \\
&=& \frac{2}{\lambda_{l}^{4}-\lambda_{m}^{4}}
(C_{l}^{\prime\prime\prime} C_{m}^{\prime\prime}-
C_{m}^{\prime\prime\prime}
C_{l}^{\prime\prime})\Bigg|_{y=\frac{1}{2}}  \qquad\text{when } l \neq m\\
&=& \frac{1}{\lambda_{m}^{4}}(\frac{1}{2}
C_{m}^{\prime\prime\prime}  C_{m}^{\prime\prime}-\frac{1}{4}(
C_{m}^{\prime\prime\prime})^2)\Bigg|_{y=\frac{1}{2}}  \qquad\text{when } l
= m ,
\label{Xdefine}
\end{eqnarray}
where $C_{m}^{\prime\prime}=D^2 C_{m}(y)$, {\it etc}. The matrix
elements $F_{lm}(\kappa, \gamma)$ were evaluated numerically for each
electrode configuration using Romberg integration\cite{recipes}. The
divergences in $Q(y)$ at the edges of the film are overcome because
$C_l(y)$ goes to zero sufficiently fast at $y= \pm 1/2$. The
functions $\Theta_{m}$ and ${\Omega}_{sm}$ are simple smooth functions
for the first few values of $m$ and are straightforward to integrate
numerically.

\subsection{Marginal Stability}

\begin{figure}
\epsfxsize =6in
\centerline{\epsffile{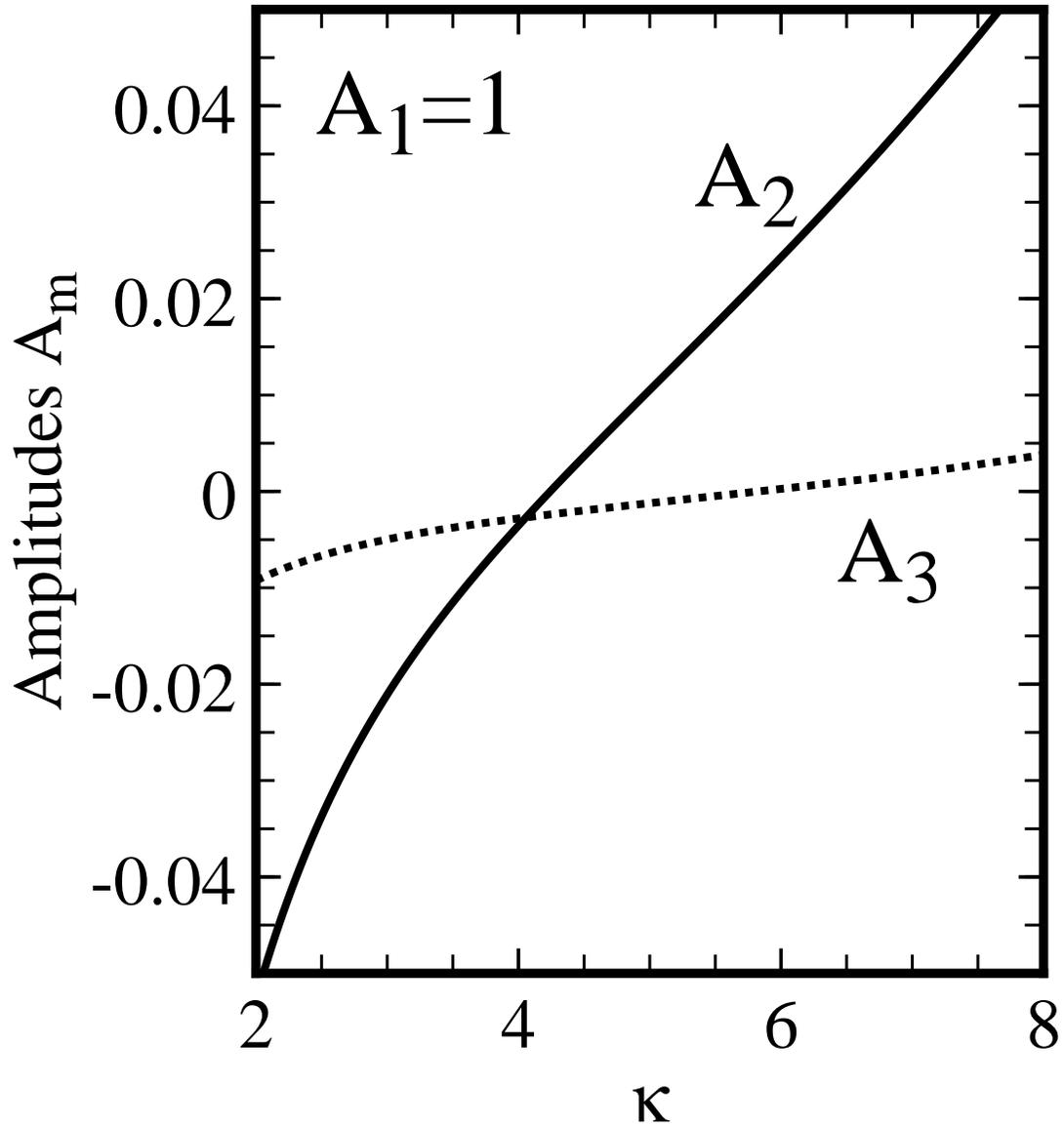}}
\vskip 0.1in
\caption{The amplitudes $A_2$ and $A_3$ of the second and third
terms of the expansion for $\Lambda(y)$, Eq.
(\protect\ref{lamdaexpandchan}).  }
\label{ampplot}
\end{figure}

To find the conditions for marginal stability, we set the growth rate
of the perturbations $\gamma$ equal to zero in the compatibility
condition, Eq. (\ref{compat}).  The Prandtl-like dimensionless group
${\cal P}$ drops
out, so that the marginal stability conditions are independent of ${\cal P}$,
just as is the case in the B\'enard problem.  Eq.
(\ref{compat}) then implicitly defines the marginal stability curve
${\cal R}={\cal R}_o(\kappa)$. We proceeded as follows. Choosing a
value of $\kappa$, we set $l=m=1$ and calculate $F_{11}(\kappa)$.
Then Eq. (\ref{compat}) can be simply solved to get the first
approximation ${\cal R}_o^{[1]}(\kappa)$. We then find
$F_{lm}(\kappa)$ for $l,m=1,2$ and search near ${\cal
R}_o^{[1]}(\kappa)$ for roots of the $2 \times 2$ determinant, Eq.
(\ref{compat}),  to find ${\cal R}_o^{[2]}(\kappa)$. We can then use
$A_1=1$ to find $A_2$ in Eq. (\ref{lamdaexpandchan}) by
backsubstitution. We carried this algorithm to the third order in the
Chandrasekhar expansion, for which the maximum value of $|A_3|$ is of
order $10^{-2}$ and the resulting neutral curve ${\cal R}_o(\kappa)$
no longer changes significantly. Fig. \ref{ampplot} shows the
amplitudes $A_2$ and $A_3$ for the wire case, relative to $A_1=1$.  It
is clear that the higher terms in the Chandrasekhar expansion
contribute very little to the sum in Eq. (\ref{lamdaexpandchan}).

\begin{figure}
\epsfxsize =6in
\centerline{\epsffile{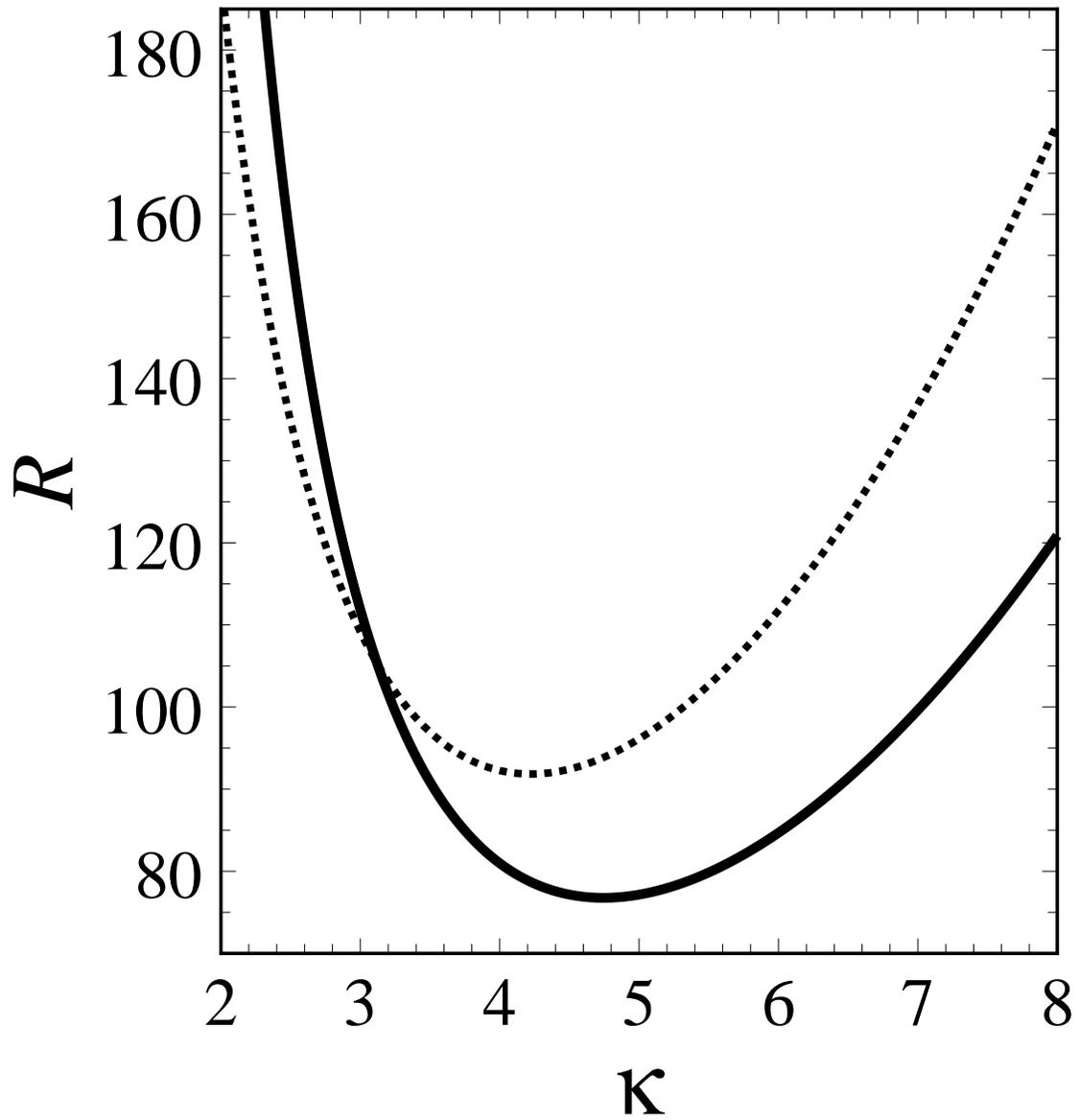}}
\vskip 0.1in
\caption{The neutral curve
for plate (dashed line) and wire (solid line) electrodes.}
\label{Rvskappa}
\end{figure}

Fig. \ref{Rvskappa} shows the neutral curve for the plate and wire
electrode configurations.  The minima of these curves define the
critical values $\kappa_c$ and ${\cal R}_c = {\cal R}_o(\kappa_c)$ for
each case. These values are listed in Table \ref{rkxitable}.  We find
that both neutral curves give $\kappa_c$ between $4$ and $5$, but that
${\cal R}_c$ is lower for the wire electrode case. This is apparently
due to the steeper slope of $q^{(0)}(y)$, evident in Fig. \ref{qvsy},
for the case of wire electrodes. Neither value of $\kappa_c$ is
particularly close to the B\'enard value of 3.117, but they are in
reasonable agreement with the value determined from the smectic film
experiments\cite{oldprl,PRApaper,glepaper}, as discussed in section
\ref{discussion} below.

We can define a length scale $\xi_0$ in terms of the curvature of
${\cal R}_o(\kappa)$ near $\kappa_c$
\cite{revmodphys,ahlersreview,marco}:
\begin{equation}
\xi_0^2 = \frac{1}{2}\frac{d^2 \epsilon_c}{d
\kappa^2}\Bigg|_{\kappa=\kappa_c},
\label{xi0define}
\end{equation}
where $\epsilon_c = ({\cal R}_o(\kappa)/{\cal R}_c)-1$. This length
will appear as a coefficient in an amplitude equation description of
the convection pattern near onset\cite{revmodphys,ahlersreview,marco}.
To find $\xi_0$ accurately, we fit
$\epsilon_c$ to a parabola over a range $\kappa=\kappa_c \pm
\Delta \kappa$ and then systematically reduced $\Delta \kappa$ until
the value of $\xi_0$ taken from the fit became independent of $\Delta
\kappa$. This corresponded to a fitting range $\epsilon_c
\leq 5 \times 10^{-4}$. The values of $\xi_0$, given in Table
\ref{rkxitable}, were slightly dependent on the electrode
configuration.

\subsection{The Linear Growth Rate \lowercase{$\gamma$}}

Returning to the full compatibility condition Eq. (\ref{compat}) with
$\gamma \neq 0$, we consider the behavior of the growth rate
$\gamma$ of the linear modes near the critical values of ${\cal R}$
and $\kappa$. The time scale $\tau_0$, defined by
\begin{equation}
\tau_0^{-1}  =  \frac{\partial \gamma(\epsilon)}{\partial
\epsilon}\Bigg|_{\kappa=\kappa_c, \epsilon=0 } ,
\label{tau0define}
\end{equation}
where $\epsilon = ({\cal R}/{\cal R}_c)-1$, will also appear in an
amplitude equation description of the pattern near onset
\cite{revmodphys,ahlersreview,marco}.

\begin{figure}
\epsfxsize =6in
\centerline{\epsffile{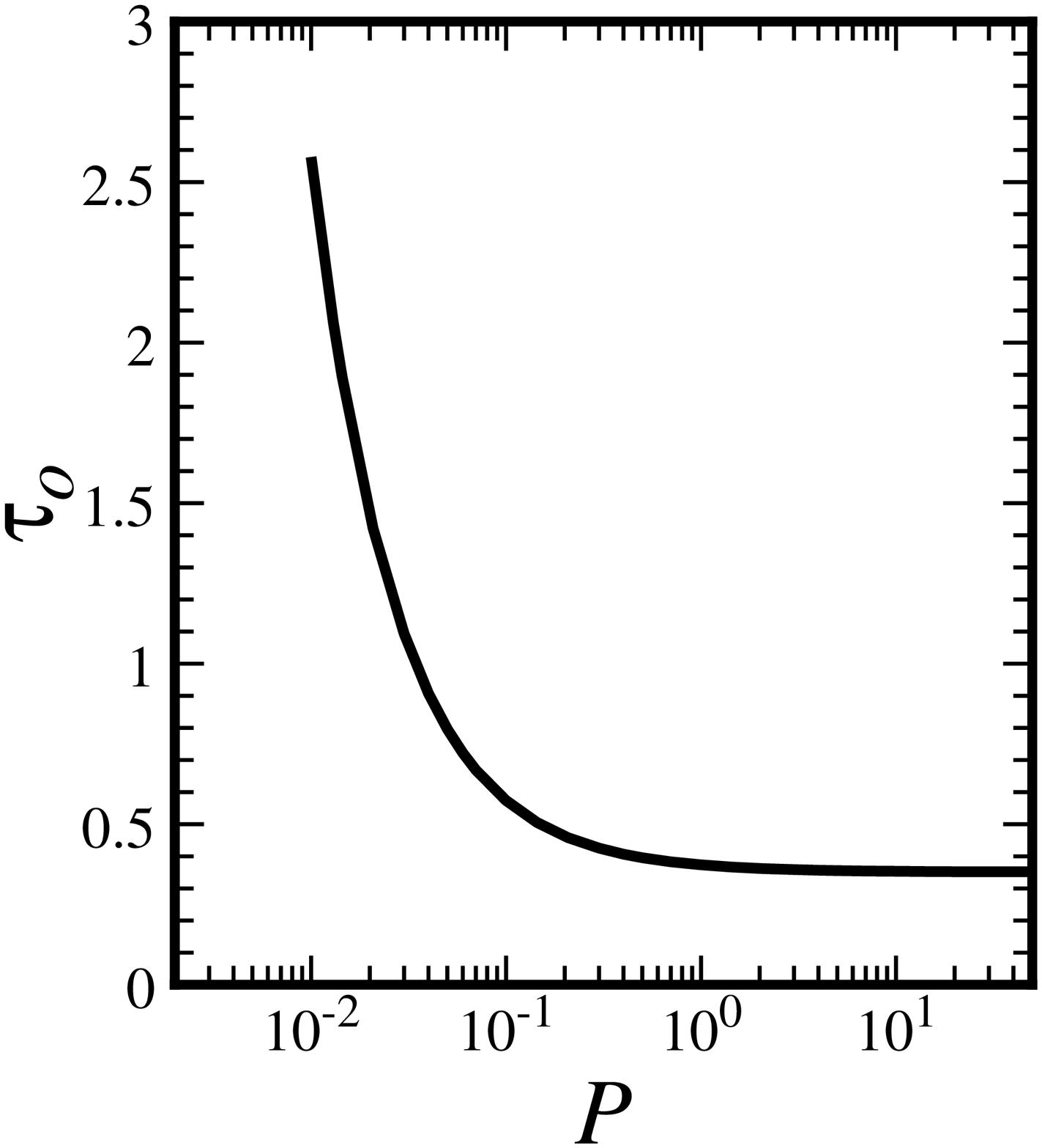}}
\vskip 0.1in
\caption{$\tau_0$ as a function of ${\cal P}$ for wire electrodes.}
\label{tauvsp}
\end{figure}

The matrix element $F_{lm}(\kappa,\gamma)$ is rather expensive to
calculate for $\gamma \neq 0$, because we must use the iteration
scheme outlined in Section \ref{gammanonzero}. It is most
computationally efficient to choose a value of $\gamma$, fix
$\kappa=\kappa_c$, and then solve Eq. (\ref{compat}) for ${\cal R}$.
This was done for ten values of $\gamma$ in the range $-0.1 \leq
\gamma \leq +0.1$, using three Chandrasekhar modes. The results
depend on ${\cal P}$. The resulting function $\gamma(\epsilon)$ is
very nearly linear in $\epsilon$ with a ${\cal P}$-dependent slope and
$\gamma(0)=0$.  We determined $\tau_0$ from polynomial fits to
$\gamma(\epsilon)$ for ${\cal P} \geq 0.01$.  The results are only
slightly dependent on electrode configuration.  $\tau_0$ is plotted as
a function of ${\cal P}$ for wire electrodes in Fig. \ref{tauvsp}.
For ${\cal P} > 1$, $\tau_0$ tends towards the limiting values given
in Table \ref{rkxitable}.

\section{Discussion}
\label{discussion}

The stability analysis presented above demonstrates that a thin, weakly
conducting suspended fluid film becomes unstable to spatially periodic
convective flow if a sufficiently large voltage is applied across the
film. Since our analysis is linear, it cannot describe the convection
pattern above onset, but it does provide important information about
the onset of convection. In this Section, we discuss the theoretical
results in the light of previous experiments on smectic
\cite{oldprl,jstatphys,PRApaper,glepaper} and nematic
\cite{faetti} films.

Eq. (\ref{Rdefine}), combined with the neutral curve, predicts that the
onset of convection occurs at a critical voltage proportional to the
film thickness $s$, and independent of the film width $d$, given by
\begin{equation}
V_c = \frac{s}{\epsilon_0} \sqrt{\sigma \eta {\cal R}_c}\>.
\label{Vcdefine}
\end{equation}
The dependence of $V_c$ on $\sqrt{\sigma \eta}$ follows inevitably
from dimensional analysis. Faetti {\it et al}. also found a critical
voltage proportional to $s \sqrt{\sigma \eta}$ and independent of $d$
from a highly simplified model of the ``vortex mode" observed by them in
nematic films \cite{faetti}.

In experiments on convection in smectic films, $V_c$ has been found to
be proportional to $s$ for films up to about 20 molecular layers
(i.e., about 63 nm) thick \cite{glepaper}.  For larger $s$, $V_c$
grows somewhat more slowly. This may be a sign that layer-over-layer
shears in the $z$ direction exist for thicker films; such flows are
not accounted for in our calculation.  A linear dependence of $V_c$ on
$s$ has also been observed in experiments on nematic
films\cite{faetti}. The nematic films are much thicker than the
smectic films, and have significant thickness nonuniformities. The
also exhibit slow flows even below the onset of convection, making
$V_c(s)$ rather difficult to measure.

Our most recent experiments on smectic films \cite{glepaper} show no
dependence of $V_c$ on $d$ for films with $d$ between 0.7 and 2.0 mm,
with  thicknesses $s$ between two and 25 molecular layers, that is,
between 6.3 nm and 80 nm. This is consistent with the prediction of Eq.
(\ref{Vcdefine}). Over about the same range of thickness, as noted above,
$V_c$ is also proportional to $s$, as predicted.  A weak variation of
$V_c$ with $d$ was, however, observed in our earlier work \cite{PRApaper}
for $d$ in the larger range of 0.36 mm to 3.5 mm.  This work used a
thicker film (107 molecular layers, or 340 nm) and a slightly different
electrode configuration, with guard electrodes outside the main electrode
wires. These features may have contributed some $d$-dependent
three-dimensional effects.

The wavenumber at onset observed in smectic film experiments is
\cite{glepaper} $k_c^{expt} = 4.94 \pm 0.03~d^{-1}$. The measured
value of $d$ is uncertain to $\pm 5$\%, so this result yields a
measured dimensionless wavenumber $\kappa_c^{expt} = 4.94 \pm 0.25$.
This is in satisfactory agreement with the value of $\kappa_c = 4.74$
found from the minimum in the calculated neutral curve for wire
electrodes. At present, no data is available for comparison to the
predictions for the plate electrode geometry.

It is interesting to note that the charge relaxation time $\tau_q =
\epsilon_0 d / \sigma s$ appropriate to thin conducting films
emerges as the natural unit of time in our analysis.  As discussed
in Ref. \cite{PRApaper}, in a thin film the relaxation time is
greater than the bulk value $\epsilon/\sigma$ by the factor $ d/
\epsilon_r s$, where $\epsilon_r$ is the relative permittivity of
the fluid. This is a consequence of the restricted geometry, and the
fact that the fields lie in the free space outside
the film.  The wave number of the convection pattern observed at onset
changes when the film is driven with ac voltages for frequencies much
larger than $1/\tau_q$.  It would be interesting to modify our analysis
to allow for time-periodic driving voltages.

It is often useful to describe patterns near onset with an equation
for the slowly varying amplitude $A$ of the pattern. For one-dimensional
systems
which are symmetric under $A \rightarrow -A$, the
appropriate amplitude equation is the Ginzburg-Landau equation
\cite{revmodphys,ahlersreview},
\begin{equation}
\label{gle}
\tau_0 {\partial A \over \partial t} = \epsilon A - g |A|^2 A
+ \xi_0^2 {\partial ^2 A \over \partial x^2} \>.
\end{equation}
Here the amplitude $A$ can be taken as the amplitude of the convective
velocity field, and $g$ governs the nonlinear saturation of the
amplitude. $\xi_0$ and $\tau_0$ are characteristic length and time
scales introduced earlier.  We have previously demonstrated that
measurements near the onset of convection in smectic films can be
described well by Eq. (\ref{gle}) with $\epsilon = (V/V_c)^2 -
1$\cite{PRApaper,glepaper}. The onset of convection occurs at a
supercritical bifurcation, and the $\epsilon$ dependence of the flow
velocity above onset, the behavior of the amplitude of convection near
a lateral boundary and the relaxation of the pattern amplitude after
sudden changes in $\epsilon$ are all well described by fits to Eq.
(\ref{gle}). Eq. (\ref{gle}) can also be derived from the full
electrohydrodynamic equations of motion presented above \cite{vatche}.

Our analysis gives theoretical results for $\xi_0$ and $\tau_0$, as
discussed above. Our predicted value for correlation length is
${\xi_0} = 0.285$, which is about 20\% smaller than the experimental
value \cite{glepaper} of $\xi_0^{expt} = 0.36 \pm 0.03$. This is in
fair, but not completely satisfactory, agreement.
To arrive at both
$\kappa_c^{expt}$ and $\xi_0^{expt}$, the experimental measurements
were made nondimensional by dividing by the measured film width $d$,
which is known to within about 5\%.
The main obstacle to making
quantitative comparisons to our predictions for $V_c(s)$ and $\tau_0$
are the poorly known material parameters $\sigma$ and (especially)
$\eta$.  These appear in the expression for the slope $V_c(s)/s$, and
are also required for calculating nondimensional times. Realistic
smectic films are sufficiently viscous that they have values of ${\cal
P} \gg 1$, so that we expect the infinite-${\cal P}$ limit of $\tau_0$
to apply.  The conductivity $\sigma$ for the doped smectic liquid
crystal used in our experiments has been measured at 1 kHz in a bulk
sample\cite{gleeson} and at dc in an annular film\cite{zahirunpub}.
Over this frequency range, it changes by a factor of three.  To get
agreement between $\tau_0^{expt}$ and our theoretical value requires a
value of $\sigma$ which lies between the dc and 1 kHz measurements.
Agreement with the $V_c(s)/s$ data \cite{glepaper}, however, requires
using a value of $\eta$ a factor of 20 larger than that estimated by
extrapolating measurements of $\eta$ made in the higher temperature
nematic phase \cite{viscosity}.  This discrepancy may be a result of
neglecting the effects of air drag on the moving film, which are
likely to be important for thin, fast-moving films \cite{drag}.

The instability we have described occurs in thin films of fluids which
are isotropic in the plane of the film. It should also exist for
anisotropic fluid films near dc, for example in films of smectic C and
C$^*$ materials. In smectic C materials, the molecules are tilted with
respect to the layer planes, so the layers are two-dimensional analogs
of a nematic fluid. Smectic C$^*$ materials have an additional broken
symmetry which allows a spontaneous electric dipole moment in the
plane of the layers.  Flows in these materials will involve strong
orientational effects.  It should be straightforward to generalize our
analysis to the anisotropic case, which may lead to interesting new effects.
Recently, it has been
suggested that electroconvection, driven by the analog of the
Carr-Helfrich mechanism which operates in negative dielectric
anisotropy nematics\cite{nematicreview}, may occur in smectic C films under
ac voltages\cite{SmCpaper}.  If materials with the right parameters exist, it
seems likely that the new instabilities will coexist or compete at low
frequencies with the instability we have considered here.
Something of this sort is observed in nematic
films\cite{faetti,faettidomainmode} in
which both a ``vortex" and a ``domain" mode are found.

\section{Conclusion}
\label{conclusion}

We have presented a linear stability analysis for the onset of
electroconvection in a thin conducting fluid with two free surfaces.
We found the neutral stability curve ${\cal R}_o(\kappa)$, along with
its critical values  ${\cal R}_c$ and $\kappa_c$, and the correlation
length ${\xi_0}$ implied by its curvature near $\kappa_c$.
The linear growth rate was used to find the characteristic time
$\tau_0$. This was done for two simple electrode configurations.
These results were compared with experiments, mainly on smectic
A films.  Several generalizations of this analysis were suggested.

\acknowledgements
We would like to thank S. S. Mao and T. Molteno for interesting
discussions, and J.T. Gleeson for making conductivity measurements.
This research was funded by the Natural Sciences and Engineering
Research Council of Canada.

\vfill\eject

\begin{table}
\caption{Numerical results.}
\label{rkxitable}
\begin{tabular}{lcc}
electrode geometry&wires & plates \\ \hline
critical wavenumber, ${\kappa}_c$ & 4.744 & 4.223 \\
critical control parameter, ${\cal R}_c$ & 76.77 & 91.84 \\
correlation length, ${\xi}_0$ & 0.2843 & 0.2975 \\
time scale, $\tau_0({\cal P} = \infty)$ & 0.351 & 0.372 \\
\end{tabular}

\end{table}

\vfill\eject

\vfill\eject

\vfill\eject

\vfill\eject

\vfill\eject

\vfill\eject


\begin{references}
\bibitem{oops} There is an error in Ref. \protect{\cite{PRApaper}}: the sign is
incorrect in Eq. (B4) and $\alpha$ and $\beta$ are interchanged in Eqs. (B4)
and (B5).
\end{references}
\end{document}